\documentstyle[12pt]{article}
\begin{document}
\title{Entanglement versus Disentanglement: Quantum Cryptography} 
\author{Arindam Mitra
\\Lakurdhi, Tikarhat Road, Burdwan, 713102.\\
 West Bengal,  India.}

\maketitle
\textwidth 11 true cm
\begin{abstract}\bf
In quantum information,
the  role of entanglement and disentanglement is itself a subject
of research and debate.  Earlier works on quantum cryptography
have almost established that entanglement has no special advantage in
quantum cryptography. In this paper we reveal that entanglement
is better ingredient than disentanglement for our alternative
quantum cryptography.
\end{abstract}

\newpage
\noindent 
In quantum information, there are some tasks which can  be accomplished
only by entanglement -such as dense coding [1] and teleportation [2].
But, there are some other tasks
which can be realized both by entanglement and disentanglement.
Quantum computation algorithm [3] and quantum cryptography [3-5]are two major applications
of both entanglement and disentanglement.\\

Between entanglement and disentanglement
which one is better ingredient in quantum computation and
quantum cryptography ?
This question is yet not settled in quantum computation, however,
entanglement enjoys some favoritism from the researchers in this field.
In quantum cryptography this question is believed to have been settled.
In his  entanglement-based quantum key distribution protocol [5]
Ekert pointed out that quantum encryption can be executed
after the transmission of the quantum state.
This seemed to be advantageous to ensure the security of the key.
Bennett, Brassard and Mermin [6] comparatively studied Ekert's
entanglement-based and Bennett-Brassard's disentanglement-based
quantum key distribution (QKD) protocols.
They concluded that so far security is concerned,
the operational {\em advantage} of Ekert's protocol is  apparent.
Since then, many quantum cryptographic protocols have  been proposed
and both type of cryptosystems have been extensively studied.
But neither of the two type of systems can stake claim of its superiority.\\

In their comparative study [6], Bennett {\em et al}
made an another important observation.
They found  that  entanglement and disentanglement based cryptosystem are
indistinguishable.
That is, which type is being used cannot be distinguished by
others. If sender uses entangled state 
but tells dishonestly to the receiver that he used disentangled state
for the encryption, then receiver  could not also verify the veracity of sender's statement. In that sense, two
cryptosystems are indistinguishable. It is recently understood  conventional
quantum bit commitment protocol ( a cryptographic application)
 completely fails [7,8] because of this
indistinguishability of two systems. Therefore,  Bennett {\em et al}'s  work
has become helpful to examine other cryptographic tasks.
Their work was based on conventional cryptography.  Recently alternative
disentanglement and entanglement based cryptographic protocols
have been proposed [9,10].  Many conclusions
drawn  from conventional quantum cryptography do not hold good in
alternative quantum cryptography. So a fresh comparative study is necessary. \\

Alternative disentanglement-based cryptosystem [9] uses mixed quantum 
state to encode a bit value but alternative entanglement based system [10] 
uses many pure entangled states for
the same purpose. 
 Despite this dissimilarity, they have many similarities.
Both can operate  entirely on  quantum channel and can
 provide quantum authentication. In both the systems, key can carry
meaningful information. 
Secure bit commitment encoding [11]and
secure  quantum coin tossing [11] are possible  for both the systems.
\\

Yet the two systems are not well understood. We have seen that classical
channel cannot be used in disentanglement-based system when 
each individual bit  is
separately made secure, but we do not know whether same is true for our
entanglement-based
system.  We also do not know whether conventional cryptography or its
prototype can be recovered from these alternative systems or not.
Here we shall see that on these two questions two cryptosystems differ. \\

First, we shall present a modified (alternative)
entanglement-based QKD protocol in
which classical channel can be used when each bit is separately made secure
and
a prototype of conventional QKD protocol can be recovered from this protocol.
This kind of modification is not possible for our disentanglement based
system. This will  imply that
our entanglement-based system can be made much much faster than
our  disentanglement-based system.

Suppose a source emits pairs of spin $1/2$ particle in their singlet state.
Two users, Alice and Bob, get one particle from  each pair.
Alice
and Bob secretly share the information of two sequences of measurements.
Suppose two sequences of direction of spin-measurements are:\\
$ S_{0}^{n} = \left\{x, \, x, \, y,
 \, y, \, x, \, y, \, y, \, y,
 \, x, \, x, \, y, \, x,..............\right\}$, \\
$ S_{1}^{n} = \left\{y, \, x, \, x, \, x, 
\, 
y, \, y, \, x, \, y, \, 
y, \, y, \, x, \, x,..............\right\}$, \\
where 0 and 1 in the subscripts 
stand for bit values and "x" and "y" are 
two orthogonal directions of measurements. \\

Let us assume they jointly decide the bit values. The bit values can be decided
when both of them use the same sequence of measurements. To produce a key,
both use $S_{0}$ and $S_{1}$ at random on their own sequences of EPR
particles. When both use $S_{0}$ or $S_{1}$, the corresponding results
will be perfectly correlated. But if one use $S_{0}$ and other $S_ {1}$
or vice versa, the results will not be perfectly correlated. So $50\%$
bit value choices are discarded.  The remaining $50\%$ bits
form the key. We shall first assume they
 reveal results through classical public channel.  \\

Their  measurements yield
the two sequences of  data sets:\\

\noindent 
$\left\{R_{1}^{A}, \, R_{2}^{A},\, R_{3}^{A},\, R_{4}^{A},\, R_{5}^{A},\,
R_{6}^{A},\,R_{7}^{A},\,R_{8}^{A},................\right\}$.\\ 
$\left\{R_{1}^{B}, \, R_{2}^{B},\, R_{3}^{B},\, R_{4}^{B},\, R_{5}^{B},\,
R_{6}^{B},\,R_{7}^{B},\,R_{8}^{B},................\right\}$. \\The first one
is Alice's sequence and second one is Bob's.
Half of their data sets contain 
perfectly correlated
data.\\

Eavesdropper's problem is to know the secret code of measurements.
For simplicity, let us think they want to produce a single bit and
 only Bob's  particles are exposed to
Eve. Eve can directly or  indirectly measure using her own
sequence of measurements. She gets a set of data from her measurements
and taps Alice's set of data when Alice reveals the results.
But  these two sets of data will neither reveal  any bit information
nor complete information of Alice's choice of measurements.
Now it is Eve's turn to reveal the results. Alice's and Eve's  
results cannot be perfectly 
correlated. But this can be interpreted by Alice as a case of 
non-identical choice of bit values. Note that in $S_{0}$ and $S_{1}$ there
is a common subsequence $S_{c}$. So if Alice does not get perfect
correlation, she can check the data corresponding to $S_{c}$. Irrespective
of choice of sequences of measurements, the  data  corresponding to
$S_{c}$ will be always perfectly correlated. This second test will expose 
eavesdropping. Still it is not the last nail to eavesdropping. \\

The data are  not secure because public channel is not authenticated channel.
Eve can impersonate. After Alice's disclosure of data, Eve, impersonating 
Bob, can reveal "fake data" correlating with Alice's data. Same thing she
can do with Bob's data impersonating  Alice. Note that this attack
works only for "fake correlation". That is, this attack will work when the users
choose the same bit value. But  they also choose different bit values
 in $50\%$ cases. In those cases,  data are not perfectly correlated, 
only the data
corresponding to $S_{c}$ are perfectly correlated. So initially
if they do not get perfect correlation between their
data sets, they will get perfect correlation in the  subsets. As
$S_{c}$ is hidden in $S_{0}$ and $S_{1}$, Eve could not generate 
"fake correlation" in the data corresponding to the  subset $S_{c}$.
Therefore Eve can only impersonate to select the bits not to reject  the 
bits.  Eve can leave the task of rejecting the bits for
 the legitimate users. It seems that system fails. \\

There is a rescue.  The "fake correlation" attack works
as both of them reveal all the data of the same events. The "fake correlation"
cannot be produced if they do not reveal any data. But the data has to
be revealed if the system is to run. If they do not reveal all the results
of the same events, yet the system can work but "fake correlation" attack
cannot. For clarity, suppose they divide the results of each set
into two subsets. Alice's subsets are $r_{1}^{A}$ and $r_{2}^{A}$ and
Bob's corresponding subsets are:  $r_{1}^{B}$ and $r_{2}^{B}$. They 
reveal the data of non identical sets -that is, either
$r_{1}^{A}$ and $r_{2}^{B}$ or  $r_{2}^{A}$ and $r_{1}^{B}$.
Because, the data of two correlated subsets  are not revealed,
 "fake correlation" attack will not
work. \\

To create many bits, the strategy, discussed above, has to
be repeatedly used to ensure  bit by bit security.
If any bit is found corrupted, the next bit will not be produced.
If eavesdropping is detected they 
must reject $S_{0}$  and $S_{1}$ and may try with another two 
preselected sequences of measurements. \\

 Is it possible to recover existing quantum cryptography  from alternative
 quantum cryptography and vice versa ?  
 Let us see  the basic difference of
 the conventional and alternative systems. In conventional quantum cryptography, a
  pure state or a pure  entangled state represent a  classical bit/bits.
  On the other hand
 in alternative quantum cryptography, many states represent a  classical bit.
 The bits of the conventional cryptosystem do not carry   meaningful
 information but it  carries meaningful information in alternative
 cryptosystem. Therefore recovery of alternative system from conventional system
 is not possible. But if we can produce  pure state -bits (which 
  does not carry any
 meaningful information) from  alternative system then at least
 recovery of prototype of conventional system, if not the same system,
 will be possible. We have two options - recovery of conventional entanglement-based
 system from alternative entanglement based system and conventional
 disentanglement-based system from alternative disentanglement-based system.
 Next we shall see that the  former can be easily realized.
 \\

We have
seen that when both of them use $S_{0}$ or $S_{1}$, the data are perfectly correlated.
These two sets of data can make a key provided they are not revealed. Suppose Alice
divides the results into three subsets $r_{1}^{A}$, $r_{2}^{A}$ and $r_{3}^{A}$.
Taking the results of same instances( which events will be chosen to construct
the three subsets are not secret) Bob prepares his three subsets
$r_{1}^{B}$, $r_{2}^{B}$ and $r_{3}^{B}$.
Alice reveals $r_{1}^{A}$ and Bob $r_{2}^{B}$ or Alice reveals $r_{2}^{A}$
and  Bob  $r_{2}^{B}$. Both go through the correlation test. 
If correlation is
 found they know their chosen bit value and side by side they know the undisclosed
  subsets $r_{3}^{A}$ and $r_{3}^{B}$ contain perfectly correlated data.
 To construct $r_{3}$, it is better to use the data corresponding 
to $S_{c}$
 so that they always get perfectly correlated data even when they use
 non-identical  sequences of measurements. Continuing the process two different
 kind of  keys (fast and slow keys) can be produced. 
The former does not exist even in the mind of the
 users and the later can exist in the mind of the users. If they want to produce only the former type
 they can share only a single sequence of measurement instead of the two.\\

The recovery of conventional entanglement-based system from alternative
entanglement-based system is
not possible. The reason is, a sequence of single photon polarized states produces
sequence of results. These results can represent bit value. But if
these results are revealed they cannot be secure. \\

Why do these systems differ on the above two issues ?
From a  sequence of two-particle entangled state Alice and Bob
can get correlated random bits. And some of the correlated data
are used for authentication via public channel and rest of the correlated
data itself can make a fast key (which is a recovery of a prototype of
conventional QKD protocol). Measurements on a
sequence of disentangled states never produce correlated data.  Therefore,
entanglement is a necessary condition to have the above mentioned
two utilities. As we get fast key from  entanglement based system, entanglement
is better secure
number generator than  disentanglement in our case. \\

Throughout our discussion we relied on bit by bit security. If we
do not want  bit by bit security but want security of many bits (meaningful)
at a time, then such security needs to be proved. In that scenario there
may have a possibility of using authenticated classical channel (authentication by some
additional shared classical bits) in our
disentanglement-based system.  In that eventuality, we can say the use of classical
channel and  bit by bit security are mutually exclusive in our disentanglement-based
system. \\


\begin{thebibliography}{99}  
\bibitem{bb92}C.H. Bennett and S. Wiesner, Phy. Rev. Lett, {\bf 69}, 21, (1992).
\bibitem{bb93}C.H. Bennett, G. Brassard, C. Crepeau, R. Jozsa, A. Peres and W. K. Wootters.
{\bf 70}, 195(1993).
\bibitem{ch20}C. H. Bennett, D. P. Divincenzo. Nature, {\bf 404}, 247, (2000).
\bibitem{bb84}C.H. Bennett and G. Brassard, {\em in Proceedings of
IEEE International Conference on Computers, System and Signal Processing, Bangalore, India,
1984 } (IEEE, NewYork, 1984), pp. 175-179.
\bibitem{ek91}A. Ekert, Phys. Rev.  Lett. {\bf
67}, 661 (1991). 
\bibitem{bbm92}C. H. Bennett, G. Brassard, and N. D. Mermin,  
 Phys.
Rev. Lett, {\bf  68}, 557 (1992).  
\bibitem{ma97}D. Mayers,  Phy. Rev. Lett, {\bf 78}, 3414  (1997).
\bibitem{lo97}H -K. Lo and C. F. Chau,  Phy. Rev. Lett,
{\bf 78}, 3410, (1997).
\bibitem{ari98}A. Mitra. http://xxx.lanl.gov/quant-ph/9812087 (submitted to PRL).
\bibitem {ari99}A. Mitra. http:// xxx.lanl.gov/physics/0007074 (Submitted to PRL).
\bibitem{ar20}A. Mitra.http://xxx.lanl.gov/physics/0007089 (submitted to PRL).

\end{thebibliography}
\end{document}